\documentclass[10pt,journal,compsoc]{IEEEtran}
\usepackage{cite}
\usepackage[cmex10]{amsmath}
\usepackage{graphicx,epsfig,amssymb}
\usepackage{anysize}
\usepackage{graphicx,cite,epsfig,subfigure}
\usepackage{amsmath,amssymb,geometry}
\usepackage{amsmath,bm}
\usepackage{setspace}
\usepackage{multirow}
\usepackage{booktabs}
\usepackage{amsmath}
\usepackage{xcolor}
\usepackage{algorithm} 
\usepackage{algorithmic} 
\usepackage{textcomp} 
\usepackage{float} 

\begin{document}
\title{{Entropy-Based Energy Dissipation Analysis of Mobile Communication Systems}}

\author{Litao~Yan,~\IEEEmembership{Student Member,~IEEE,}
       Xiaohu~Ge,~\IEEEmembership{Senior Member,~IEEE,}

\thanks {Litao Yan, Xiaohu Ge (corresponding author) are with School of Electronic Information and Communications and National International Joint Research Center
of Green Communications and Networking (No. 2015B01008), Huazhong University of Science and Technology, Wuhan 430074, Hubei, P. R. China.
Email: \{litaoyan, xhge\}@hust.edu.cn}
}

\thispagestyle{empty} \maketitle \thispagestyle{empty}

\setcounter{page}{1}\begin{abstract}
Compared with the energy efficiency of conventional mobile communication systems, the energy efficiency of fifth generation (5G) communication systems has been improved more than 30 times. However, the energy consumption of 5G  communication systems is 3 times of the energy consumption of fourth generation (4G) communication systems when the wireless traffic is increased more than 100 times in the last decade. It is anticipated that the traffic of future sixth generation (6G) communication systems will keep an exponential growth in the next decade. It is a key issue how much space is left for improving of energy efficiency in mobile communication systems. To answer the question, an entropy-based energy dissipation model based on nonequilibrium thermodynamics is first proposed for mobile communication systems. Moreover, the theoretical minimal energy dissipation limits are derived for typical modulations in mobile communication systems. Simulation results show that the practical energy dissipation of information processing and information transmission is three and seven orders of magnitude away from the theoretical minimal energy dissipation limits in mobile communication systems, respectively. These results provide some guidelines for energy efficiency optimization in future mobile communication systems.
\end{abstract}


\newpage
\section{Introduction}
With the global deployment of the fifth-generation (5G) communication systems, the mobile traffic is increasing rapidly in an exponential trend. In parallel with this growth, the energy consumption of mobile communication systems is also increasing in an urgent way \cite{ge2017}. Moreover, the massive information processing of base station obviously increases the computation power of mobile communication systems. However, traditional energy consumption studies can not catch up with the requirements of energy efficiency optimization for mobile communication systems. Recent studies on energy dissipation was investigated for the ultra-low energy consumption computation systems \cite{conte2019}. Nevertheless, the studies about the energy dissipation is rarely mentioned for mobile communication systems. Therefore, it is time to explore a new way for analyzing the energy dissipation of mobile communication systems.

For a long time, to reduce the energy consumption of communication systems, the most commonly used method is to design algorithms to effectively allocate resources of the system. Along this way, the energy efficiency is improved at the cost of a reduced communication capacity. For instance, a resource scheduling strategy based on a self-organization rule was proposed to minimize the transmit energy consumption of femtocell networks \cite{lopez2013}. For MIMO (multiple-input and multiple-output) communication systems, an online optimization strategy was proposed to maximize the energy efficiency for dynamic scenarios \cite{merti2016}. With the development of massive MIMO and millimeter wave technologies, the hybrid precoding was proposed to reduce energy consumption and system cost \cite{heath2016}. Taking the computation power into consideration, an alternating optimization algorithm was proposed to jointly optimize the communication power and computation power of massive MIMO communication systems \cite{ge2018}. More recently, reconfigurable intelligent surface (RIS) is emerging as a new solution to improve the energy efficiency of wireless communication systems \cite{zhang2020}. To maximize the energy efficiency under the constraint of the user transmission rate, a joint optimization algorithm was proposed for transmit beamforming and RIS control \cite{yangz2022}. It is worth pointing out that the goal of almost all the above studies is to maximize the ratio of spectral efficiency to the overall energy consumption, which is currently the most widely accepted measure of energy efficiency, where the overall energy consumption is usually defined by the sum of the transmit power and other hardware energy consumption \cite{miao2010}. However, this definition is not appropriate from a physical perspective because the transmit power can be recovered in some way \cite{visser2013} and it should be excluded from the energy efficiency formula. Moreover, the same question is exist in the measurement of hardware energy consumption. The developments of simultaneous wireless information and power transfer (SWIPT) and wireless energy harvesting are also an example to defend the above idea \cite{perera2017,wagih2021}. Therefore, what we should optimize is the energy which is really wasted in mobile communication systems. Generally, the wasted energy of mobile communication systems is dissipated into the environment as the heat \cite{zhirnov2014}.

	The link between the energy dissipation and information has been studied for decades in the field of thermodynamics. The lower bound on energy dissipation for irreversible logical operations was first discussed rigorously by Landauer \cite{landauer1961}. Landauer pointed out that any logical state corresponds to a physical state, and irreversible logic processes reduce the degree of freedom of the physical state, which inevitably leads to an energy dissipation. This conclusion is also known as Landauer's principle and has been experimentally verified \cite{berut2012}. Luca et. al, analyzed the current semiconductor technology and found that the switching energy consumption of transistors ($10^{-18}$J) is three orders of magnitude away from the theoretical limit Landauer gave \cite{luca2015}. Based on significant breakthroughs in non-equilibrium statistical physics \cite{crooks, jarzynski}, the research on the energy and information in real-world systems emerged and became more extensive and in-depth. Landauer's principle was extended to analog computing systems by establishing the relationship between information erasure and continuous phase transitions \cite{diamantini2016}. Relating information theory and stochastic thermodynamics in nonlinear dynamical systems, the thermodynamic cost of modularization when erasing information was derived based on the ratchet model \cite{boyd1}. In addition, the impact of the rate of information processing and the scale of the processor on the required computation energy consumption was analyzed \cite{boyd2}. For realistic computers, the relationship between the reliability of computation and energy dissipation was also studied \cite{boyd3}. By analyzing the thermodynamic cost of three main methods of generating random numbers, Aghamohammadi, et. al, pointed out that the thermodynamic cost of creating information can complement the theory proposed by Landauer about the physics of erasing information \cite{aghamohammadi2017}. More recently, a new computing paradigm, i.e., the thermodynamic computing, which refers to computations under the scale between classical and quantum computing, was discussed in detail \cite{conte2019}.

	Although the relationship between information and energy is being extensively studied in the field of physics, there are very few studies which apply these theoretical results for practical mobile communication systems. As a matter of fact, researchers from the communication industry usually consider the information and energy as two independent objects for optimizing separately. Moreover, there is a lack of systematic study on the communication architecture from the perspective of energy dissipation in the open literatures, which makes it challenging to guide the design of future low-power mobile communication systems. Considering the above problems, we propose an entropy-based energy dissipation model for mobile communication systems for the first time. The main contributions and innovations of this paper are summarized as follows:
\begin{enumerate}
\item Based on the architecture of Carnot heat engine, an energy dissipation architecture including the information processing and information transmission is proposed for mobile communication systems, where the energy input and dissipation are coupled with the information input and output.
\item Under the proposed energy dissipation architecture, the energy dissipation of information processing and information transmission is analyzed using tools from non-equilibrium statistical physics for mobile communication systems. For the information processing with the logical irreversible calculation, the lower bound of energy dissipation of the logic gate circuit is derived. Moreover, the energy dissipation models of three typical modulations, i.e., the channel estimation, linear processing and channel coding are obtained. Based on the thermodynamic principle of the measurement-erasure cycle, the energy dissipation models of five typical modulations, i.e., the amplification, filtering, analog-to-digital conversion, frequency mixing and phase shifting are derived for the information transmission. Furthermore, the theoretical minimal energy dissipation limits are derived for typical modulations in mobile communication systems.
\item Based on the proposed energy dissipation models, simulation results show that the energy dissipation increases with the system bandwidth, while the relationship between the number of communication system users and energy dissipation is non-monotonic. In addition, simulation results indicate that there exists a gap of three orders of magnitude between the current technology and the theoretical limit in the energy dissipation of information processing. In comparison, the gap between the current technology and the theoretical limit in information transmission is more than seven orders of magnitude.
\end{enumerate}
	
The remainder of the article is organized as follows. Section II introduces some necessary basics of thermodynamics. In Section III, an energy dissipation architecture of mobile communication systems is proposed. Section IV and Section V describe energy dissipation models of typical modulations for information processing and information transmission of mobile communication systems, respectively. In Section VI, the energy dissipation model is simulated and the comparisons with the traditional energy consumption model are analyzed. Finally, Section VII concludes the whole paper.

\section{Thermodynamic Preliminaries}

We use $x$ to denote a specific physical state of a physical system. For a system composed of multiple particles, $x$ might be a multi-variable vector, where each component of the vector may itself be a vector, denoting physical quantities associated with the particle, e.g., its charge, spin, and so on. The state space $\mathcal{X}$ is the set of all possible physical states of the system. Depending on different scenarios, $\mathcal{X}$ can be either mesoscopic or microscopic, corresponding to varying degrees of coarse-graining of the information-carrying physical space \cite{kolchinsky2020}. Because of the uncertainty of the system, the distribution of physical states is used to describe a system. When a system undergoes a physical process with an initial state distribution ${{p}^{\text{ini}}}$ and an end state distribution ${{p}^{\text{end}}}$, we use the following equation to describe the process
\[{{p}^{\text{end}}}({{x}^{\text{end}}})=\sum\limits_{x}{P({{x}^{\text{end}}}|x){{p}^{\text{ini}}}(x)},\tag{1}\]
where ${{x}^{\text{end}}}$ is the end state of the system, $P({{x}^{\text{end}}}|x)$ is the conditional probability of the end state ${{x}^{\text{end}}}$ given the initial state $x$. In continuous time, the above equation is rewritten as
\[\frac{d}{dt}{{p}_{t}}({x}')=\sum\limits_{x}{{{p}_{t}}(x){{K}_{t}}(x\to {x}')},\tag{2}\]
where ${{p}_{t}}({x}')$ is the probability when the state of system is ${x}'$ at time $t$, and ${{K}_{t}}(x\to {x}')$ is the state transition rate matrix from the state $x$ to the state ${x}'$. Using (2), the \emph{entropy flow} of the system is obtained as
\[{{S}_{F}}=\sum\limits_{x,y}{{{p}_{t}}(x){{K}_{t}}(x\to {x}')\ln \frac{{{K}_{t}}(x\to {x}')}{{{K}_{t}}({x}'\to x)}}.\tag{3}\]
where ${{S}_{F}}$ is the thermodynamic entropy flows from the system to the environment.

The thermodynamic entropy $S(p)$ of the system with state distribution $p$ is defined through the Shannon entropy $H(p)$ \cite{parrondo2015}
\[S(p)=kH(p)=-k\sum\limits_{x}{p(x)\log p(x)},\tag{4}\]
where $k$ is Boltzmann’s constant, $p(x)$ is the probability of the system in state $x$.

The \emph{entropy change} $\Delta S$ of the system is defined as the difference of the thermodynamic entropy between the initial and the end state of a physical system, that is
\[\Delta S=S({{p}^{\text{ini}}})-S({{p}^{\text{end}}}).\tag{5}\]

The \emph{entropy production} ${{S}_{G}}$ is the increase in the total thermodynamic entropy of the system and the thermal reservoir after the physical process. It is a non-negative physical quantity which measures the irreversibility of the physical process \cite{seifert2012}. Specifically, the entropy production is zero when a physical process is carried out in a thermodynamically reversible way.

The relationship between the entropy flow, the entropy change of the system and entropy production satisfies the entropy balance equation \cite{seifert2012}, i.e.,
\[{{S}_{F}}=\Delta S+{{S}_{G}}.\tag{6}\]
In the case where the local detailed balance \cite{peliti2021} holds, the entropy flow is equal to the energy dissipation   of the system when the temperature is normalized, i.e.,
\[Q=T{{S}_{F}},\tag{7}\]
where $T$ is the absolute temperature.

As an example of the application of the entropy balance equation, we now analyze the energy dissipation of a simple information erasure process. Suppose there is a system with only two states in its state space, corresponding to the logic value of 0 and 1, respectively. Assume that the distribution of the initial state of the system is uniform and the end state is fixed and known. Therefore, the entropy change of the system after the erasure process is calculated as $\Delta S=k(\log 2-0)=k\log 2$. If the process is carried out in a thermodynamically reversible way, i.e., the entropy production ${{S}_{G}}=0$. Then according to (6) and (7), it is easy to obtain the energy dissipation of the erasure process $Q=kT\log 2$, which is the lower bound on the energy dissipation of 1-bit information erasure, also known as Landauer limit \cite{bennett2003}.

\section{Energy Dissipation Architecture for Mobile Communication Systems}
To achieve a complete communication process, a mobile communication system needs to receive the information as the input and output their own information. From the viewpoint of energy, mobile communication systems take the energy as input and convert it into output radio signals while generating the energy dissipation, i.e., the energy wasted in the form of heat. For ideal mobile communication systems with no energy dissipation, the spectral efficiency is identical to the Carnot efficiency\cite{ge2022}. Inspired by the architecture of Carnot heat engine, in this paper, the physical processes of information operations are divided into information transmission and information processing, corresponding to different physical characteristics. Based on the division, the energy dissipation architecture for mobile communication systems is shown in Fig. 1, where functional modules not directly related to the information, such as the control and active cooling modules, are ignored.
\begin{figure}
\includegraphics[width=3in]{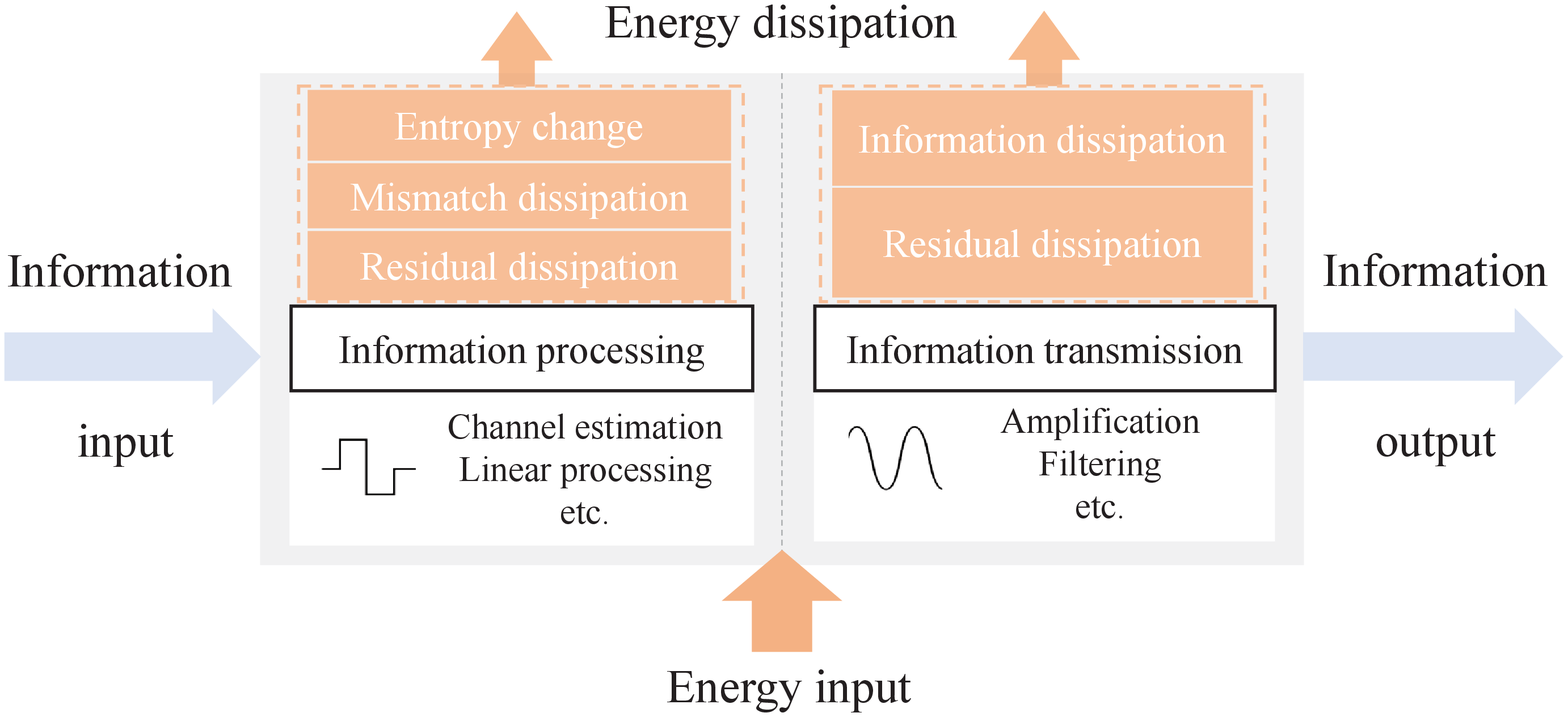}
\begin{quote}
\small Fig. 1. Energy dissipation architecture of mobile communication systems.
\end{quote}
\end{figure}
\subsection{Information Processing}
In mobile communication systems, the information processing refers to the realization of specific operations on information through logical operations, and the objects of operations are digital signals. The physical realization unit of an information processing device is the logic gate. During the logic operation, the phase transition of the logic gate is discrete, that is, its input/output state space is limited. The logic gate returns a value based on a specific mapping rule. The mapping is logically irreversible when the input state space is larger than the output state space. Therefore, when the logic gate undergoes an irreversible process, it will lose some information. Based on Landauer’s principle, the energy dissipation is inevitably generated. As shown in Fig. 1, in this paper, the energy dissipation of information processing is further divided into three types: (1) the dissipation caused by the entropy change, (2) the dissipation caused by the mismatch between the actual input distribution and the optimal distribution of logic gates, (3) the dissipation independent of the input, which is also called the residual dissipation. The calculation of the dissipation will be described in detail in Section IV. Typical information processing modulations include channel estimation, linear processing, and channel coding for mobile communication systems.
\subsection{Information Transmission}
Unlike the information processing, the purpose of information transmission is to preserve the integrity of the signal carrying the information as much as possible. The information transmission is realized by analog circuits made up of fundamental electronic elements such as resistance, capacitance, etc. The objects to be operated are analog signals, and their phase transitions are continuous, meaning there are infinitely many states during the operation. In the real world, the information transmission takes place within a finite time and is thermodynamically irreversible, so there will also be inevitable energy dissipation. As shown in Fig. 1, in this paper, the energy dissipation of information transmission is further classified into two types: (1) the information-related dissipation, which is called the information dissipation, (2) the information-independent residual dissipation. Section V introduces the detailed calculation of information transmission energy dissipation. Typical information transmission modulations include amplification, filtering, analog-to-digital conversion, frequency mixing, and phase shifting for mobile communication systems.

Based on different characteristics of the information processing and information transmission, we will discuss the energy dissipation generated by these two fundamental physical processes in mobile communication systems.

\section{Energy Dissipation of Information Processing}
Considering an arbitrary logic gate as the physical system to be analyzed, the physical process performed by the system is the logic computation. The initial state of the system is the input of the logic gate and the end state is the output of the gate. For a logic gate with the initial state space $\mathcal{X}$ and end state space $\mathcal{Y}$, the conditional probability $p(y|x)$ is used to describe a specific computation of the logic gate, where $x\in \mathcal{X}$ and $y\in \mathcal{Y}$. For the initial states $x,\text{ }{x}'$ and an end state $y$, if $p(y|x)>0,\text{ }p(y|{x}')>0$  and ${x}'\ne x$, we define that $x$ and ${x}'$ belong to the same state set, denoted by $\mathcal{L}$. The initial state space is denoted by $\mathcal{X}={{\mathcal{L}}_{1}}\cup {{\mathcal{L}}_{2}}\cup \cdots \cup {{\mathcal{L}}_{n}}$. The conditional probability of the occurrence of state $x$ in ${{\mathcal{L}}_{i}}$ is written as
\[{{p}^{{{\mathcal{L}}_{i}}}}(x)=\frac{p(x)}{{{p}_{{{\mathcal{L}}_{i}}}}},\tag{8}\]
where ${{p}_{{{\mathcal{L}}_{i}}}}=\sum\limits_{x\in {{\mathcal{L}}_{i}}}{p(x)}$ is the probability that the system state belongs to the state set ${{\mathcal{L}}_{i}}$. For example, considering a NAND gate with the initial state space ${{\mathcal{X}}_{\text{NAND}}}=\{(0,0),(0,1),(1,0),(1,1)\}$ and the end state space ${{\mathcal{Y}}_{\text{NAND}}}=\{0,1\}$, the state transition probability matrix ${{P}_{\text{NAND}}}$ is
\[{{P}_{\text{NAND}}}=\begin{bmatrix}
   0 & 0 & 0 & 1  \\
   1 & 1 & 1 & 0
\end{bmatrix}.\tag{9}\]
The rows and columns of the matrix correspond to the initial state space and the end state space of the NAND gate, respectively. The matrix elements are the probabilities of occurrence of the corresponding state transfers. Fig. 2 depicts a schematic of the state set partitioning for the physical process of the NAND gate, where the blue and green boxes represent the initial state space divided into two state sets.
\begin{figure}[!h]
\includegraphics[width=3in]{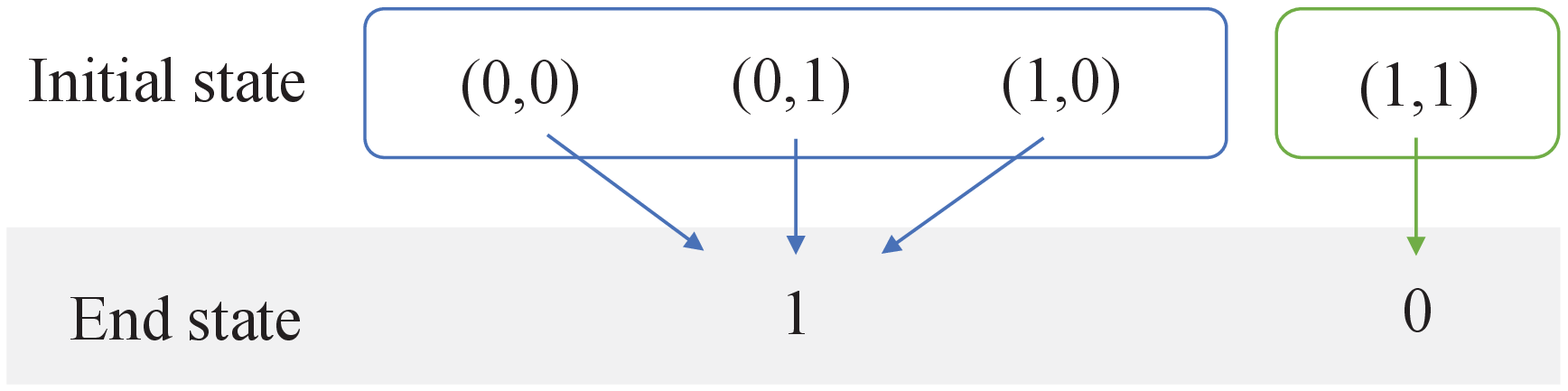}
\begin{quote}
\small Fig. 2. State set partitioning for the physical process of a NAND gate.
\end{quote}
\end{figure}

The partitioning of the physical process into state sets can provide many advantages for the thermodynamic analysis of the process, the most important of which is that each initial state set is assumed to be corresponds to an isolated sub-process and that the thermodynamic properties of the individual sub-processes are independent of each other. Under this assumption, the authors in \cite{wolpert2020} proved that for any physical process, the entropy flow of the system has the following form
\[ {{S}_{F}}=\Delta S+D({{p}_{\text{ini}}}||{{q}_{\text{ini}}})-D({{p}_{\text{end}}}||{{q}_{\text{end}}}) \\+\sum\limits_{{{\mathcal{L}}_{i}}}{{{s}_{G}}({{p}_{{{\mathcal{L}}_{i}}}})},\tag{10} \]
where ${{q}_{\text{ini}}}$ and ${{q}_{\text{end}}}$ are the optimal distribution of the initial and end state distributions, $D(\cdot)$ is the relative entropy operation, $D({{p}_{\text{ini}}}||{{q}_{\text{ini}}})$ is the relative entropy between the optimal and actual distributions of initial states, $D({{p}_{\text{end}}}||{{q}_{\text{end}}})$ is the relative entropy between the optimal and actual distributions of end states, and ${{s}_{G}}({{p}_{{{\mathcal{L}}_{i}}}})$ is the residual entropy production of the state set ${{\mathcal{L}}_{i}}$, relating only to the partition of the system state sets.

A digital circuit can be defined by a tuple $(G,P,W)$, where $G$ is the set of all logic gates in the circuit, $P$ is the set of the conditional probability of state transitions, and $W$ is the set of connecting structures of all the gates. Ignoring the feedback and cycling among logic gates, when a digital circuit is performing a logic operation, the logic gate only affects states of logic gates which are directly connected to it. In other words, a digital circuit can be divided into many independent sub-circuits, where each sub-circuit performs an independent physical process during the information processing. Take a three-input majority logic circuit as an example in this paper. The circuit consists of two AND gates, i.e., AND1 and AND2, and two XOR gates, i.e., XOR1 and XOR2, as shown in Fig. 3. The dashed lines with arrows in Fig. 3 indicate three inputs of the circuit. It takes four steps for the circuit to complete an information processing, where each step corresponds to the working of a specific logic gate, as indicated by the red box in Fig. 3. Furthermore, the circuit can be divided into two sub-circuit ${{G}_{1}}$ and ${{G}_{2}}$, where ${{G}_{1}}$ is composed by gate AND1 and XOR1, and ${{G}_{2}}$ is composed by gate AND2 and XOR2.
\begin{figure}[!h]
\centering
\includegraphics[width=3in]{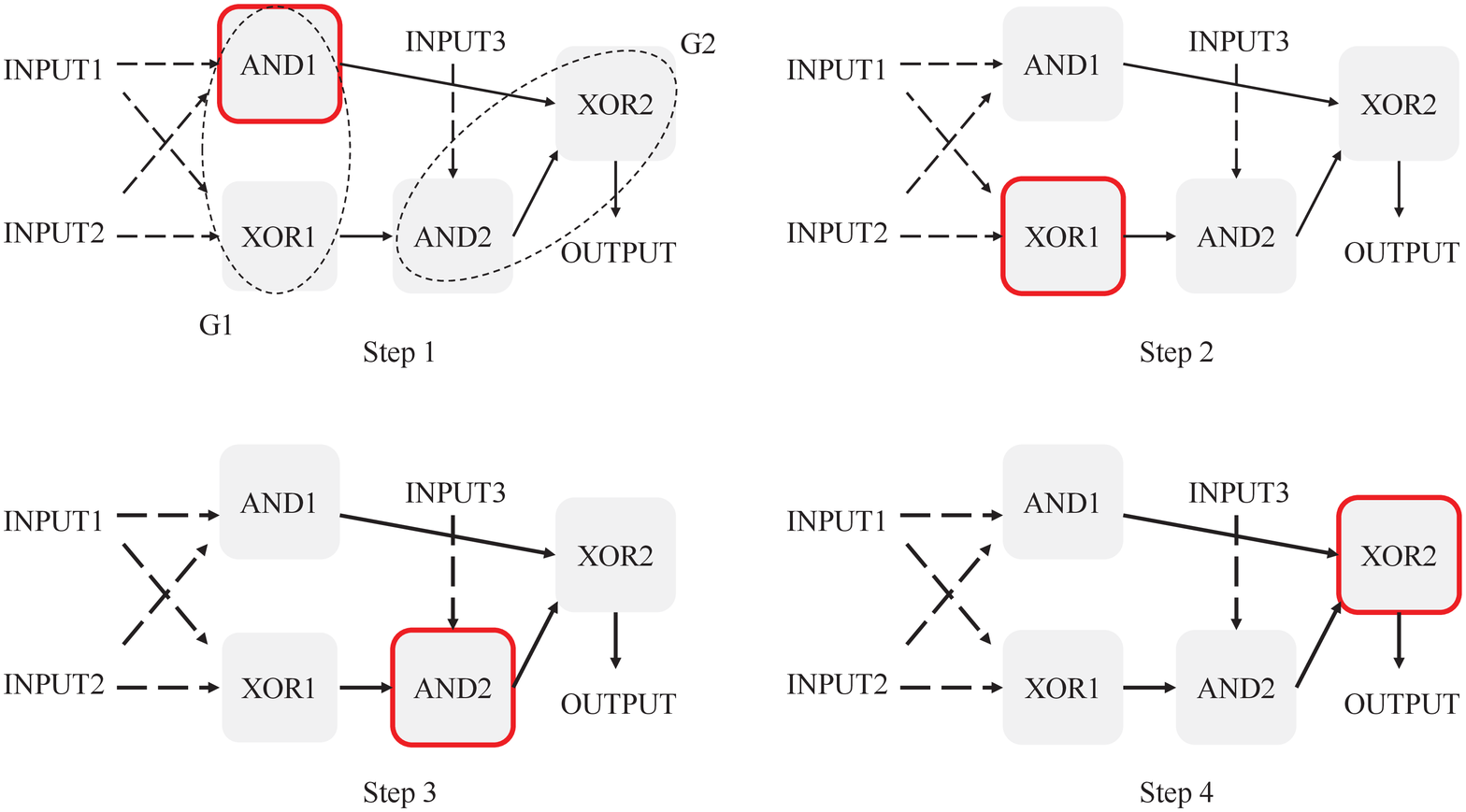}
\begin{quote}
\small Fig. 3. Physical processes of the logic computation of a majority logic circuit, where 1and 2 steps take place in ${{G}_{1}}$, 3 and 4 steps take place in ${{G}_{2}}$.
\end{quote}
\end{figure}

Due to the independence assumption of the circuit, the total entropy flow of the entire circuit ${{S}_{F}}({{G}_{1}}\cup {{G}_{2}})$ is simply the sum of the entropy flow of two sub-circuits, i.e.,
\[{{S}_{F}}({{G}_{1}}\cup {{G}_{2}})={{S}_{F}}({{G}_{1}})+{{S}_{F}}({{G}_{2}}),\tag{11}\]
where ${{S}_{F}}({{G}_{1}})$ and ${{S}_{F}}({{G}_{2}})$ are the entropy flow of sub-circuit ${{G}_{1}}$ and ${{G}_{2}}$, respectively. If we regard each logic gate $g$ as a sub-circuit, the total entropy flow of the whole circuit during the logic computation becomes a direct extension of (10), which can be written in a simplified form below
\[{{S}_{F}}=\Delta S+S_{G}^{M}+S_{G}^{R},\tag{12}\]
where the first term $\Delta S$ represents the energy dissipation caused by the change of the initial and end states of all logic gates in the digital circuit. Moreover, $\Delta S$ is expressed by the entropy change in the entropy balance equation, i.e.,
\[\Delta S=\sum\limits_{g}{\left[ S(p_{g}^{\text{ini}})-S(p_{g}^{\text{end}}) \right]},\tag{13}\]
where $S(\cdot)$ is the entropy calculation operation, $p_{g}^{\text{ini}}$ and $p_{g}^{\text{end}}$ represent the initial state distribution and the end state distribution of the logical gate $g$ in a logical operation, respectively. The second term of (12) represents the additional thermodynamic cost due to the mismatch between the actual input distribution and the inherent optimal distribution of the logic gate, which is called the mismatch dissipation. Moreover, $S_{G}^{M}$ is a kind of the entropy production in the entropy balance equation, denoted as
\[S_{G}^{M}=\sum\limits_{g}{\left[ D(p_{g}^{\text{ini}}||q_{g}^{\text{ini}})-D(p_{g}^{\text{end}}||q_{g}^{\text{end}}) \right]}.\tag{14}\]
The third term of (12) represents the dissipation generated even if the logic gate has the optimal input distribution, which is called the residual dissipation. Moreover, $S_{G}^{R}$ is also denoted by the entropy production in the entropy balance equation, i.e.,
\[S_{G}^{R}=\sum\limits_{g}{\sum\limits_{{{\mathcal{L}}_{i}}}{{{s}_{G}}({{p}_{{{\mathcal{L}}_{i}}}}(g))}},\tag{15}\]
where ${{p}_{{{\mathcal{L}}_{i}}}}(g)$ is the probability that the state of $g$ belongs to ${{\mathcal{L}}_{i}}$.

To sum up, the entropy flow of the digital circuit is expressed by
\[\begin{aligned} 
{{S}_{F}} & =\sum\limits_{g}{\left[ S(p_{g}^{\text{ini}})-S(p_{g}^{\text{end}}) \right]} \\
& +\sum\limits_{g}{\left[ D(p_{g}^{\text{ini}}||q_{g}^{\text{ini}})-D(p_{g}^{\text{end}}||q_{g}^{\text{end}}) \right]} \\
& +\sum\limits_{g}{\sum\limits_{{{\mathcal{L}}_{i}}}{{{s}_{G}}({{p}_{{{\mathcal{L}}_{i}}}}(g))}}. 
\end{aligned} \tag{16}\]

From the above analysis, it can be seen that the entropy change and mismatch dissipation are non-linearly related to the initial state distribution of the digital circuit. Therefore, it is unrealistic to directly use (16) to rigorously calculate the energy dissipation of the information processing for digital circuits which are comprised of more than ten billion transistors. On the other hand, digital circuits with arbitrary functions can be realized by a combination of NAND gates \cite{bird2014}. Taking advantage of this fact, we now simplify (16) to analyze the lower bound for the energy dissipation of a realistic digital circuit.

\emph{Theorem 1 (Lower bound of the energy dissipation):} For a digital circuit composed by $N$ NAND gates, if the number of floating-point operations (FLOs) performed by the circuit is $O$, the lower bound of the energy dissipation during the information processing is
	\[Q_{\min }^{O}=\frac{3}{4}\left( 2+\log \frac{3}{4} \right)kTON.\tag{17}\]

\emph{Proof:} Suppose the initial state and end state of a NAND gate in a logical operation are $p_{\text{NAND}}^{\text{ini}}$ and $p_{\text{NAND}}^{\text{end}}$, respectively, the entropy change of the gate is
\[\Delta S(\text{NAND})=S(p_{\text{NAND}}^{\text{ini}})-S(p_{\text{NAND}}^{\text{end}}).\tag{18}\]

Assuming that the optimal initial distribution is uniform, and the logic computation is carried out in a way which the residual dissipation is zero, then the entropy production is solely the mismatch dissipation, i.e.,
\[\begin{aligned}
   {{S}_{G}}(\text{NAND})&=D(p_{\text{NAND}}^{\text{ini}}||q_{\text{NAND}}^{\text{ini}})-D(p_{\text{NAND}}^{\text{end}}||q_{\text{NAND}}^{\text{end}}) \\
 & =\sum\limits_{p_{\text{NAND}}^{\text{ini}}}{p_{\text{NAND}}^{\text{ini}}\log \frac{p_{\text{NAND}}^{\text{ini}}}{q_{\text{NAND}}^{\text{ini}}}} \\ 
& -\sum\limits_{{{p}_{1}}}{p_{\text{NAND}}^{\text{end}}\log \frac{p_{\text{NAND}}^{\text{end}}}{q_{\text{NAND}}^{\text{end}}}} \\
 & =2-S(p_{\text{NAND}}^{\text{end}})-\left[ -S(p_{\text{NAND}}^{\text{end}}) \right. \\
& \left. +2p_{\text{NAND}}^{\text{end}}(0)-p_{\text{NAND}}^{\text{end}}(1)\log \frac{3}{4} \right].
\end{aligned} \tag{19}\]

Thus, the entropy flow of the NAND gate
\[\begin{aligned}
 {{S}_{F}}(\text{NAND}) & =\Delta S(\text{NAND})+{{S}_{G}}(\text{NAND}) \\
 & =2-2p_{\text{NAND}}^{\text{end}}(0)+p_{\text{NAND}}^{\text{end}}(1)\log \frac{3}{4} \\
 & =\left( 2+\log \frac{3}{4} \right)p_{\text{NAND}}^{\text{end}}(1).
\end{aligned} \tag{20}\]

It is obvious that the entropy flow of a NAND gate is only related to the probability of its output being 1. If the initial state distribution of a digital circuit is uniform, i.e., the probability that the input of a NAND gate is either 1 or 0 is $\frac{1}{2}$, then the probability that each NAND gate has an output of 1 is $p_{\text{NAND}}^{\text{end}}(1)=\frac{3}{4}$. Therefore, during a single FLO, the entropy flow of the circuit is $\frac{3}{4}\left( 2+\log \frac{3}{4} \right)N$, and the energy dissipation is $\frac{3}{4}\left( 2+\log \frac{3}{4} \right)kTN$. Multiplying it with the number of FLOs, (17) is obtained.

In summary, the energy dissipation of a digital circuit can be divided into two parts. One is the energy dissipation caused by the FLOs, denoted as ${{Q}^{O}}$, which depends on the specific information content. The other is the residual dissipation caused by the circuit operating in the actual thermodynamically irreversible way, denoted as ${{Q}^{R}}$, which only depends on the physical realization of the circuit and can be regarded as a constant. Therefore, the total energy dissipation of the information processing in a mobile communication system can be written as
\[Q={{Q}^{O}}+{{Q}^{R}}.\tag{21}\]

In the rest of this section, the energy dissipation of channel estimation, linear processing, and channel coding is discussed in detail based on Theorem 1.

\subsection{Channel Estimation}
In the following analysis of the energy dissipation of the information processing in mobile communication systems, a single-cell multi-user Massive MIMO system operating over a bandwidth of $B$ Hz is configured. The base station is assumed to be equipped with $M$ antennas to communicate with $K$ single-antenna users using the time-division duplex (TDD) scheme. The coherence time and coherence bandwidth of the flat-fading channel are denoted as ${{T}_{c}}$ and ${{B}_{c}}$, respectively. The channel matrix $\mathbf{H}=\left[ \mathbf{{h}_{1}},\mathbf{{h}_{2}},\ldots ,\mathbf{{h}_{K}} \right]\in {{\mathbb{C}}^{M\times K}}$, where $\mathbf{{h}_{k}}={{\left[ {{h}_{k,1}},{{h}_{k,2}},...,{{h}_{k,M}} \right]}^{T}}\in {{\mathbb{C}}^{M\times 1}}$ is the channel vector between the $k$th user and the base station antennas.

To ensure the orthogonality between pilot sequences, it is assumed that the pilots occupy $K$ symbols in the uplink \cite{mohammed2014}. Hence, the pilot sequences of $K$ users can be represented by a $K\times K$ unitary matrix $\mathbf{\Phi}$ , where each row of $\mathbf{\Phi}$ corresponds to a pilot sequence sent by a user. Ignoring the noise of antennas, the pilot sequence received by the base station is simply denoted as $\mathbf{{Y}_{p}}=\mathbf{H}\mathbf{\Phi}$. Under the minimum mean square error criterion, the estimated channel matrix is
\[\mathbf{\hat{H}}={\mathbf{{Y}_{p}}\mathbf{\Phi}^{H}}.\tag{22}\]
It can be seen from (22) that, to accomplish a channel estimation, there are inevitable operations of the multiplication of an $M\times K$ matrix and a $K\times K$ matrix. Based on the results in \cite{boyd2004}, the number of FLOs required to complete this operation is $MK(2K-1)$, including $M{{K}^{2}}$ complex multiplications and $MK(K-1)$ complex additions. However, in practical physical implementations, a single complex multiplication requires four real multiplications and two real additions, and a single complex addition requires two real additions \cite{golub2013}. Hence, the minimum number of FLOs required for a channel estimation is $6M{{K}^{2}}+2MK(K-1)$. Since a channel estimation is required in each coherence block and the number of coherent blocks per second is $\frac{B}{U}$, the number of FLOs per second during the channel estimation is
	\[{{O}_{CE}}=\frac{B}{U}\left[ 6M{{K}^{2}}+2MK(K-1) \right].\tag{23}\]
Considering the residual dissipation to be $Q_{CE}^{R}$, based on Theorem 1, the energy dissipation of channel estimation modulation ${{Q}_{CE}}$ is
\[
\begin{aligned}
{{Q}_{CE}} & =\frac{3B}{4U}\left( 2+\log \frac{3}{4} \right)\left[ 6M{{K}^{2}} \right. \\
& \left. +2MK(K-1) \right]kTN+Q_{CE}^{R}.
\end{aligned} \tag{24}\]

\subsection{Linear Processing}
To reduce the complexity and energy consumption of radio frequency (RF) systems in massive MIMO communication systems, the hybrid precoding is adopted at the base station side for information processing. In this case, the linear processing modulation includes the baseband precoding in downlinks and the signal detection in uplinks. The FLOs of linear processing modulation include the matrix multiplication of each information symbol and the calculation of the precoding matrix. Denoting the number of RF chains as ${{N}_{RF}}$, the matrix multiplication of a ${{N}_{RF}}\times K$ precoding matrix and a $K\times 1$ baseband signal vector is necessary for each precoding. To complete a precoding, the number of FLOs is ${{N}_{RF}}(2K-1)$, including ${{N}_{RF}}K$ complex multiplications and ${{N}_{RF}}(K-1)$ complex additions. The number of precoding operations per unit time $v$ is calculated by $v=\frac{R}{K}$, where $R$ is the transmission rate of the wireless communication system. To reduce the complexity, it is assumed that the simple maximum ratio transmission/combining (MRT/MRC) scheme is adopted in this paper. In this case, we only need to normalize each column of the channel matrix $H$ \cite{bjornson2015}, i.e., the normalization of $K$ $M\times 1$ vectors, and the number of complex multiplications and additions required for this operation are $2MK$ and $(M-1)K$, respectively. The precoding matrix needs to be computed in each coherence block. Therefore, the number of FLOs per second during the linear processing is
\[\begin{aligned}
{{O}_{LP}} & =\frac{R}{K}\left[ 6{{N}_{RF}}K+2{{N}_{RF}}(K-1) \right] \\
& +\frac{B}{U}\left[ 12MK+2(M-1)K \right].\end{aligned}
\tag{25}\]
Considering the residual dissipation to be $Q_{LP}^{R}$, based on Theorem 1, the energy dissipation of linear processing modulation ${{Q}_{LP}}$ is
\[\begin{aligned}
{{Q}_{LP}} & =\frac{3}{4}\left( 2+\log \frac{3}{4} \right)\left\{ \frac{R}{K}\left[ 6{{N}_{RF}}K+2{{N}_{RF}}(K-1) \right] \right. \\
& \left. +\frac{B}{U}\left[ 12MK+2(M-1)K \right] \right\}kTN+Q_{LP}^{R}.
\end{aligned} \tag{26}\]
\subsection{Channel Coding}
The function of channel coding includes the downlink coding and uplink decoding for mobile communication systems. To a large extent, the energy dissipation of the channel coding modulation depends on the chosen coding scheme. In this paper, a $(n,m)$ regular low-density parity-check (LDPC) code is considered to be used for the channel coding, i.e., $m$ information bits are transformed into code blocks with a length of $n$ for transmissions. This process is realized by multiplying the vector of information bits by the generator matrix, which involves binary multiplication and addition. For decoding processes, we consider a soft-decision iterative decoding method to ensure the error correction performance. Let the number of iterations be $l$, in each iteration, it is necessary to calculate the probability where each bit in the codeword is 0 or 1, and pass this probability information between the variable node and the check node. The inevitable operations are $n(n-2)$ multiplications of real numbers \cite{masera2005}. Compared with the energy dissipation of decoding processes, the energy dissipation of encoding processes can be negligible \cite{desset2003}. In this case, the number of FLOs performed by the channel coding operation per second is given as
	\[{{O}_{CD}}=\frac{n(n-2)lR}{m}.\tag{27}\]
Considering the residual dissipation to be $Q_{CE}^{R}$, based on Theorem 1, the energy dissipation of channel estimation modulation ${{Q}_{CD}}$ is
	\[{{Q}_{CD}}=\frac{3n(n-2)lR}{4m}\left( 2+\log \frac{3}{4} \right)kTN+Q_{CD}^{R}.\tag{28}\]
To sum up, the energy dissipation model of the information processing in mobile communication systems is written as
\[\begin{aligned}
  {{Q}_{IP}} &={{Q}_{CE}}+{{Q}_{LP}}+{{Q}_{CD}} \\
 & =\frac{3}{4}\left( 2+\log \frac{3}{4} \right)\ \left\{ \frac{B}{U}\left[ 8M{{K}^{2}}+2(6M-1)K \right] \right. \nonumber\\
 & + \frac{R}{K}\left[ 6{{N}_{RF}}K + 2{{N}_{RF}}(K-1) \right] \\
 & \left. + \frac{n(n-2)lR}{m} \right\} kTN + Q_{IP}^{R}, \nonumber
\end{aligned} \tag{29}\]
where $Q_{IP}^{R}=Q_{CE}^{R}+Q_{LP}^{R}+Q_{CD}^{R}$ is the total residual dissipation of three typical modulations for information processing, which depends on the specific physical realization of the circuit. The residual dissipation of three typical modulations has not been extended because of the limited pages in this paper.
\section{Energy Dissipation of Information Transmission}
As discussed in Section III, the information transmission is physically realized by analog circuits. To carry out a formalized thermodynamic analysis on the energy dissipation of analog circuits, various modules of an analog circuit are regarded as different sub-systems, where the input of each sub-system is the output of the previous sub-system to which it is connected. For each sub-system, the output signal is generated based on the measurement of the input signal. Therefore, the target function module is defined as the measurement system, and the previous module connected to it is defined as the measured sub-system. After the measurement, the state of the measurement sub-system evolves from the initial state ${{Y}^{\text{ini}}}$ to the end state ${{Y}^{\text{end}}}$, while the state of the measured sub-system stays as $X$, as shown in Fig. 4. At the end of a complete thermodynamic cycle, the measurement sub-system must erase the information it acquired after the measurement. That is, the state of the measurement system must experience a cycle of ${{Y}^{\text{ini}}}\to {{Y}^{\text{end}}}\to {{Y}^{\text{ini}}}$.
\begin{figure}[!h]
\centering
\includegraphics[width=3in]{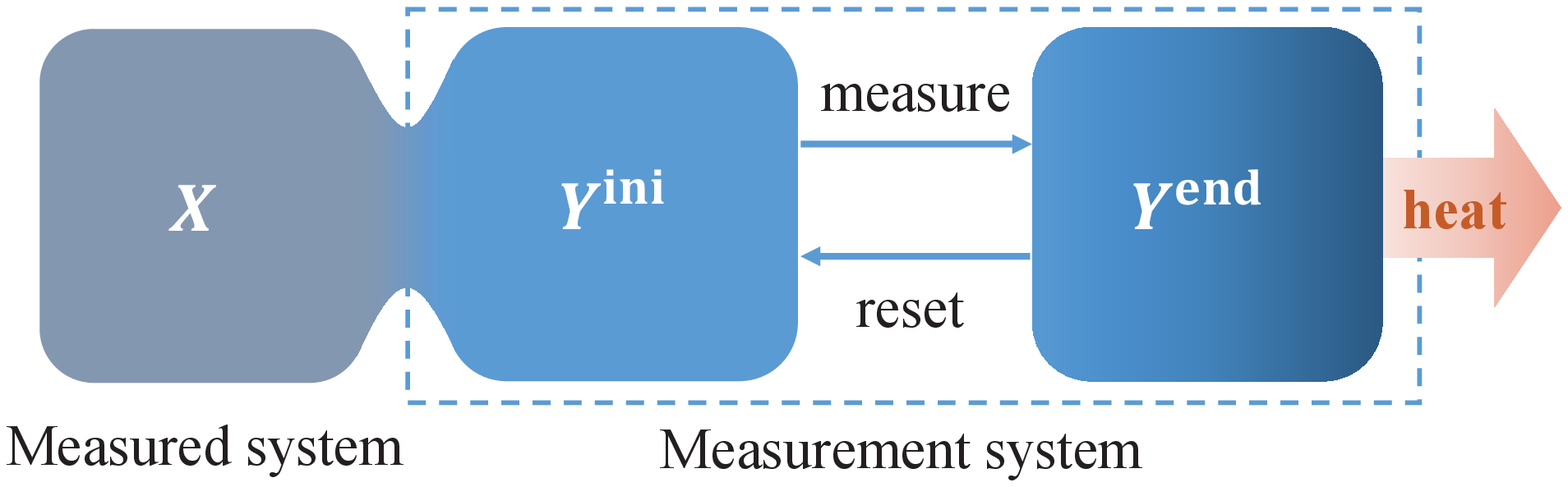}
\begin{quote}
\small Fig. 4. Thermodynamic cycle of information transmission.
\end{quote}
\end{figure}
Denote the inner energy and thermodynamic entropy of the measurement sub-system $Y$ as $E(Y)$ and $S(Y)$, respectively. The free energy of $Y$ is defined as
	\[\begin{aligned}
  & F(Y)=E(Y)-TS(Y) \\
 & =\sum\nolimits_{y}{p(y)E(y)}-\left[ \sum\nolimits_{y}{p(y)S(y)}+H(Y) \right] \\
 & =\sum\nolimits_{y}{p(y)F(y)}-kTH(Y),
\end{aligned}\tag{30}\]
where $p(y)$ is the probability that the system is in a specific state of $y$, and $F(y)$ is the free energy of the sub-system in state $y$. In the definition above, the thermodynamic entropy of the sub-system $S(Y)$ is decomposed into two parts, one is the information entropy of the sub-system $H(Y)$, the other is the average of the thermodynamic entropy of the sub-system $\sum\nolimits_{y}{p(y)S(y)}$, where $S(y)$ is called the inner entropy \cite{parrondo2015}.
Based on the second law of thermodynamics, the difference between the free energy $\Delta F$ of the initial state and the end state of a nonequilibrium system is no larger than the work done on the system, that is
	\[W-\Delta F\ge 0.\tag{31}\]
Before the measurement, the measured sub-system and the measuring sub-system are independent of each other, and the joint free energy of the system is
	\[F(X{{Y}^{\text{ini}}})=F(X)+F({{Y}^{\text{ini}}}).\tag{32}\]
After the measurement, because of the interaction between two sub-systems, the joint free energy becomes
	\[\begin{aligned}
  F(X{{Y}^{\text{end}}}) & =F(X)+\sum\nolimits_{y}{p(y)F(y)}-kTH(Y|X) \\
 & =F(X)+\sum\nolimits_{y}{p(y)F(y)}-kTH(Y) \\
& +kT\left[ H(Y)-H(Y|X) \right] \\
 & =F(X)+F({{Y}^{\text{end}}})+kTI(X;Y).
\end{aligned} \tag{33}\]
Therefore, the free energy change during the measurement is given as
	\[\begin{aligned}
   \Delta {{F}_{meas}} & =F(X{{Y}^{\text{end}}})-F(X{{Y}^{\text{ini}}}) \\
 & =F(X)+F({{Y}^{\text{end}}})+kTI(X;Y) \\
& -\left[ F(X)+F({{Y}^{\text{ini}}}) \right] \\
 & =\Delta F(Y)+kTI(X;Y),
\end{aligned} \tag{34}\]
where $\Delta F(Y)=F({{Y}^{\text{end}}})-F({{Y}^{\text{ini}}})$. The term $I(X;Y)$ is the information obtained by the measurement sub-system, i.e., the mutual information between input and output signal of the sub-system. Substituting (34) into (31), we have
	\[{{W}_{meas}}\ge \Delta {{F}_{meas}}=\Delta F(Y)+kTI(X;Y).\tag{35}\]
Based on (31), the work ${{W}_{reset}}$ required for the erasure process satisfies
	\[{{W}_{reset}}\ge \Delta {{F}_{reset}},\tag{36}\]
where $\Delta {{F}_{reset}}$ is the change of the free energy during the erasure process. Since the erasure process only occurs in the measurement system, and the measurement-erasure process forms a complete thermodynamic cycle for the measurement system. Hence, the change of the free energy for the erasure process must satisfy
	\[\Delta {{F}_{reset}}=F({{Y}^{\text{ini}}})-F({{Y}^{\text{end}}})=-\Delta F(Y).\tag{37}\]
Combined the above equations, the total work required for the analog circuit system in a measurement-erasure cycle is obtained as
	\[{{W}_{meas}}+{{W}_{reset}}\ge kTI(X;Y).\tag{38}\]
(38) clarifies the minimum energy dissipation required for any analog circuit and reveals that the minimum energy consumption of information transmission is only decided by the information obtained by the physical system. In this paper, the energy exchange between systems, i.e., external components such as fans and speakers in the circuit is ignored. Therefore, the work done on the analog circuit is ultimately converted to heat, that is
	\[{{W}_{meas}}+{{W}_{reset}}=Q.\tag{39}\]
For the convenience of analysis, $Q$ is rewritten as
	\[Q={{Q}^{I}}+{{Q}^{R}},\tag{40}\]
where ${{Q}^{I}}=kTI(X;Y)$ is the information-related dissipation of the system, which is defined as information dissipation. In contrast, ${{Q}^{R}}$ only depends on the specific physical implementation of the system, which is called residual dissipation. Specifically, in the following, five analog circuit-based information transmission modulations, i.e., the filtering, amplification, analog-to-digital conversion, frequency mixing and phase shifting modulations are analyzed.
\subsection{Filtering}
Considering a perfect analog filter circuit, the input of the circuit consists of a Gaussian signal $X$ and the thermal noise $N$. After filtering, the noise is reduced and the output is $Z=X+{N}'$. The power of $X$ is denoted by ${{P}_{fil}}$ and the power of the output noise is denoted by ${{{P}'}_{noise}}={{N}_{0}}B$, where $B$ is the bandwidth of $X$ and ${{N}_{0}}$ is the noise power spectral density. Therefore, the information obtained by the filter
\[\begin{aligned}
  & {{I}_{F}}=H(Z)-H(Z|X) \\
 & =\frac{1}{2}\log \left( 2\pi e\left( {{P}_{fil}}+{{{{P}'}}_{noise}} \right) \right) \\
 & -\frac{1}{2}\log \left( 2\pi e{{{{P}'}}_{noise}} \right) \\
 & =\frac{1}{2}\log \left( 1+\frac{{{P}_{fil}}}{{{N}_{0}}B} \right).
\end{aligned} \tag{41}\]
Since the measurement-erasure cycle is performed $B$ times per second in the filtering modulation, the information dissipation is
	\[Q_{F}^{I}=kT{{I}_{F}}B=\frac{B}{2}kT\log \left( 1+\frac{{{P}_{fil}}}{{{N}_{0}}B} \right).\tag{42}\]
When the residual dissipation is denoted as $Q_{F}^{R}$, the energy dissipation of filtering modulation ${{Q}_{F}}$ is
	\[{{Q}_{F}}=\frac{B}{2}kT\log \left( 1+\frac{{{P}_{fil}}}{{{N}_{0}}B} \right)+Q_{F}^{R}.\tag{43}\]
\subsection{Amplification}
The function of amplification is to strengthen the amplitude of the input signal so that the resolution of the output signal can be improved. For an analog amplifier circuit $G$, the input information entropy is defined as the information entropy of the input Gaussian signal $X$, that is
	\[{{H}_{A}}(X)=\frac{1}{2}\log \left( 2\pi e{{P}_{amp}} \right),\tag{44}\]
where ${{P}_{amp}}$ is the power of the input signal. The power of the output signal ${{P}_{Y}}$ is the product of the entropy power $\sigma _{g}^{2}$ of the amplifier and ${{P}_{amp}}$, i.e.,
	\[{{P}_{Y}}={{P}_{amp}}\sigma _{g}^{2}.\tag{45}\]
During the amplification, the information obtained by the amplifier is
	\[\begin{aligned}
  & {{I}_{A}}={{H}_{A}}(Y)-{{H}_{A}}(Y|X) \\
 & =\frac{1}{2}\log (2\pi e{{P}_{Y}})-\frac{1}{2}\log (2\pi e\sigma _{g}^{2}) \\
 & =\frac{1}{2}\log (2\pi e{{P}_{amp}}).
\end{aligned}\tag{46}\]
As can be seen from (46), ${{I}_{A}}$ is only decided by the input signal, which indicates that the signal can be amplified but the information cannot be increased. Thus, the information dissipation of amplification is
	\[Q_{A}^{I}=kT{{I}_{A}}B=\frac{B}{2}kT\log (2\pi e{{P}_{amp}}).\tag{47}\]
Denoting the residual dissipation as $Q_{A}^{R}$, the energy dissipation of amplification modulation ${{Q}_{A}}$ is
	\[{{Q}_{A}}=\frac{B}{2}kT\log (2\pi e{{P}_{amp}})+Q_{A}^{R}.\tag{48}\]
\subsection{Analog-to-Digital Conversion}
The analog-to-digital conversion is responsible for the quantization of the signal. When the input signal $X$ is assumed to be governed by the Gaussian distribution, the information entropy of the output signal ${{X}^{\Delta }}$ is
	\[{{H}_{C}}({{X}^{\Delta }})={{H}_{C}}(X)-{{H}_{C}}(D),\tag{49}\]
where ${{H}_{C}}(X)$ is the information entropy of $X$, and ${{H}_{C}}(D)=\log \Delta$ is called the error entropy which depends on the intervals of the length $\Delta$ when dividing $X$ \cite{kong2019}. For example, the intervals of a $b$ bit quantization precision signal is given by $\Delta ={{2}^{-b}}$. The information obtained by the analog-to-digital converter is
	\[\begin{aligned}
  & {{I}_{C}}=H(X)-H(X|{{X}^{\Delta }}) \\
 & =H(X)-H(D) \\
 & =\frac{1}{2}\log \frac{{{P}_{con}}}{{{P}_{D}}},
\end{aligned}\tag{50}\]
where ${{P}_{con}}$ is the power of the input signal, ${{P}_{D}}$ is the power of quantization noise. Based on the result of \cite{sripad}, the power of quantization noise is expressed by ${{P}_{D}}=\frac{{{\Delta }^{2}}}{12}$. Therefore, the information dissipation of analog-digital conversion is
	\[Q_{C}^{I}=kT{{I}_{C}}B=\frac{B}{2}kT\log \left( \frac{12{{P}_{con}}}{{{\Delta }^{2}}} \right).\tag{51}\]
When the residual dissipation is denoted as $Q_{C}^{R}$, the energy dissipation of analog to digital conversion modulation ${{Q}_{C}}$ is
	\[{{Q}_{C}}=\frac{B}{2}kT\log \left( \frac{12{{P}_{con}}}{{{\Delta }^{2}}} \right)+Q_{C}^{R}.\tag{52}\]
\subsection{Frequency Mixing and Phase Shifting}
Since the information-carrying degrees of freedom have not been unchanged for the frequency mixing and phase shifting modulations, the amount of information obtained by the frequency mixing and phase shifting circuits is the same. When the input power of the frequency mixing circuit is denoted as ${{P}_{mix}}$, the information of the frequency mixing circuit is
	\[{{I}_{M}}=\frac{1}{2}\log \left( 2\pi e{{P}_{mix}} \right),\tag{53}\]
and the information dissipation is
	\[Q_{M}^{I}=kT{{I}_{M}}B=\frac{B}{2}kT\log \left( 2\pi e{{P}_{mix}} \right).\tag{54}\]
Therefore, the energy dissipation of frequency mixing modulation ${{Q}_{M}}$ is
	\[{{Q}_{M}}=\frac{B}{2}kT\log \left( 2\pi e{{P}_{mix}} \right)+Q_{M}^{R},\tag{55}\]
where $Q_{M}^{R}$ is the residual dissipation of the frequency mixing circuit.
	When the input power is denoted as ${{P}_{shf}}$ and the residual dissipation is denoted as $Q_{S}^{R}$, the energy dissipation of phase shifting circuits modulation ${{Q}_{S}}$ is
	\[{{Q}_{S}}=\frac{B}{2}kT\log \left( 2\pi e{{P}_{shf}} \right)+Q_{S}^{R}.\tag{56}\]
To sum up, the total energy dissipation of the information transmission in mobile communication systems ${{Q}_{IT}}$ is
\[\begin{aligned}
{{Q}_{IT}} & ={{Q}_{G}}+{{Q}_{F}}+{{Q}_{C}}+{{Q}_{M}}+{{Q}_{S}} \\
 & =\frac{B}{2}kT\left[ \log \left( 1+\frac{{{P}_{fil}}}{{{N}_{0}}B} \right)+\log (2\pi e{{P}_{amp}}) \right. \\
 & +\log \left( \frac{12{{P}_{con}}}{{{\Delta }^{2}}} \right)+\log \left( 2\pi e{{P}_{mix}} \right) \\
 & \left. +\log \left( 2\pi e{{P}_{shf}} \right) \right]+Q_{IT}^{R},
\end{aligned}\tag{57}\]
where $Q_{IT}^{R}=Q_{G}^{R}+Q_{F}^{R}+Q_{C}^{R}+Q_{M}^{R}+Q_{S}^{R}$ is the total residual dissipation of five typical modulations for the information transmission, which depends on the specific physical realization of the circuit. The residual dissipation of five typical modulations has not been extended in information transmission because of the limited pages in this paper.

\section{Numerical Results and Analysis}
To analyze the impact of energy dissipation on the performance of mobile communication systems, relationships between the energy dissipation and performance indicators such as bandwidth and the number of users are simulated. Specifically, a mobile communication system with millimeter-wave massive MIMO and LDPC codes is configured for simulations. The equivalent zero-forcing (EZF) precoding \cite{yang2022} is used to maximize the spectral efficiency. Other simulation parameters are shown in Table I. Since the residual dissipation is decided by specific physical implementations, the energy dissipation simulated in this paper is merely considered to be the information dissipation, which also represents the theoretical lower limit of the energy dissipation.
\begin{table}[!htb]
\centering
\caption{\ Simulation Parameters }
\begin{tabular}{c|c}
    \toprule
    \textbf{Parameters } & \textbf{Value} \\
    \hline
    Transmit power ${{P}_{\text{T}}}$ & 5 W  \\
    Number of antennas $M$ & $256$ \\
    Channel coherence bandwidth $B_{c}$ & 100 MHz \\
    Channel coherence time $T_{c}$ & 35 \textmu s\\
    The number of multi-paths & 8\\
    Number of RF chains ${{N}_{\text{RF}}}$ & 32\\
    Noise power spectral density &-174 dBm/Hz\\
    Path loss exponent & 4.6\\
    Entropy power of amplifier & 5\\
    Number of transistors $N$ & ${{10}^{8}}$\\
    Quantization precision & 8 bit \\
    Decoding iterations & 10\\
        \bottomrule
        \end{tabular}
\label{tab1}
\end{table}

When the number of users is configured as 20, the energy dissipation of the information processing with respect to the bandwidth is simulated in Fig. 5. Based on the results of Fig. 5, the energy dissipation of modulations of linear processing, channel estimation and channel coding increases with the increase of bandwidth. When the bandwidth is fixed, the energy dissipation of the linear processing modulation is larger than the energy dissipation of the channel estimation modulation, and the energy dissipation of the channel estimation modulation is larger than the energy dissipation of the channel coding modulation.

Fig. 6 shows the energy dissipation of the information processing with respect to the number of users. The energy dissipation of the linear processing and channel coding modulations decreases with the increase of the number of users, while the energy dissipation of the channel estimation modulation increases with the increase of the number of users. As a result, the energy dissipation of the information processing decreases with the increase of the number of users when the number of users is less than or equal to 19. The energy dissipation of the information processing increases with the increase of the number of users when the number of users is larger than 19.
\begin{figure}[!htb]
\centering
\includegraphics[width=3in]{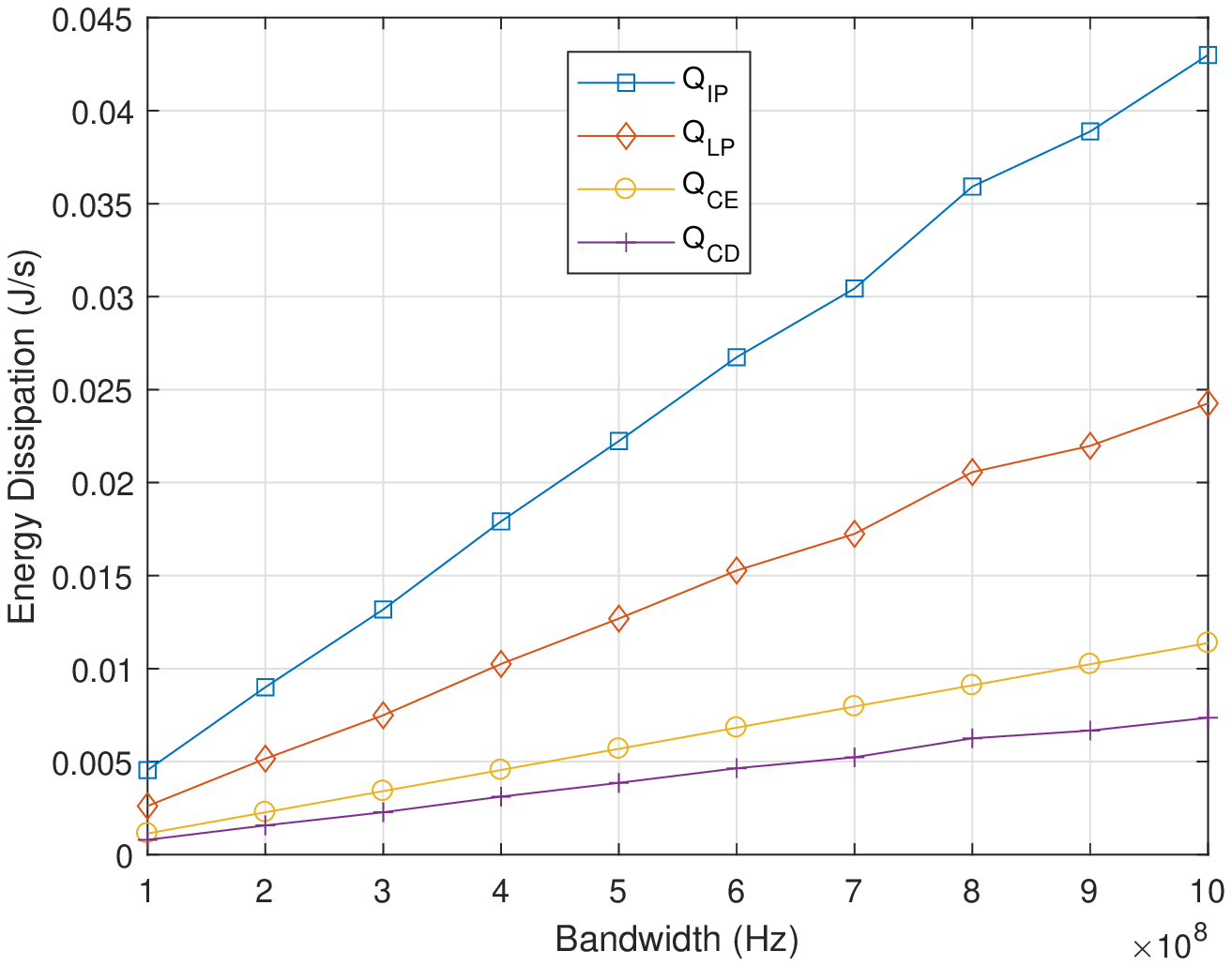}
\begin{quote}
\small Fig. 5. Energy dissipation of the information processing with respect to bandwidth.
\end{quote}
\end{figure}
\begin{figure}[!htb]
\centering
\includegraphics[width=3in]{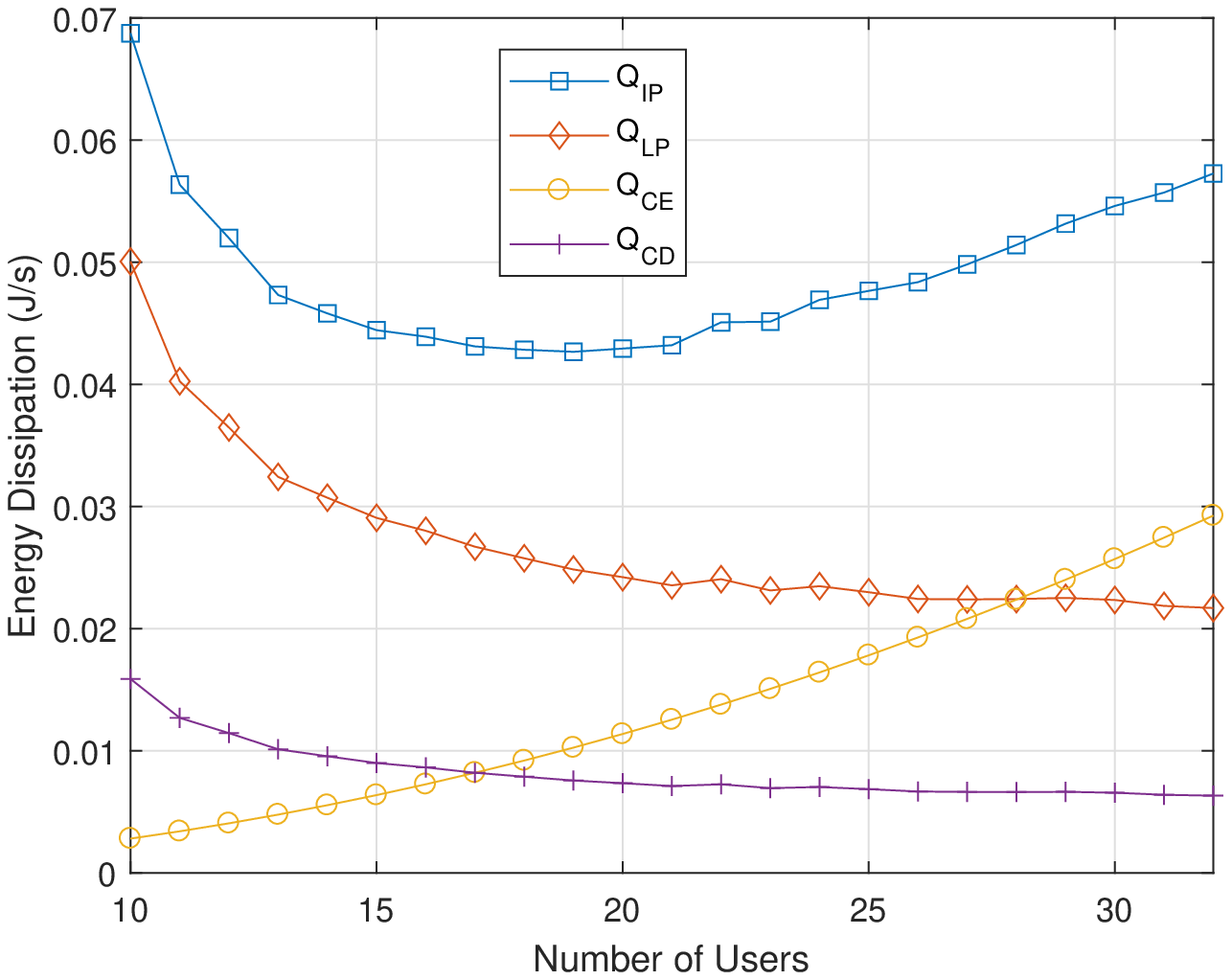}
\begin{quote}
\small Fig. 6. Energy dissipation of the information processing with respect to the number of users.
\end{quote}
\end{figure}

In the simulation of the energy dissipation of the information transmission, the circuit structure in \cite{yang2019} is adopted, where the input power of different modulations satisfies ${{P}_{amp}}={{P}_{con}}={{P}_{mix}}={{P}_{shf}}$ and ${{P}_{fil}}={{P}_{T}}$. The number of filters is configured as the same as the number of antennas $M$, and the number of phase shifters ${{N}_{shf}}$ satisfies ${{N}_{shf}}=M{{N}_{RF}}$. The energy dissipation during a single measurement-erasure cycle for different modulation circuits is shown in Fig. 7. Based on the results of Fig. 7, the energy dissipation of the filtering modulation is the largest in the measurement-erasure cycle. In contrast with the traditional impression that the amplifier consumes most of the energy, the energy dissipation of the amplification modulation is the minimum in Fig. 7.  The reason is that the information loss is the minimum for the physical process of amplification in information transmissions.
\begin{figure}[!htb]
\centering
\includegraphics[width=3in]{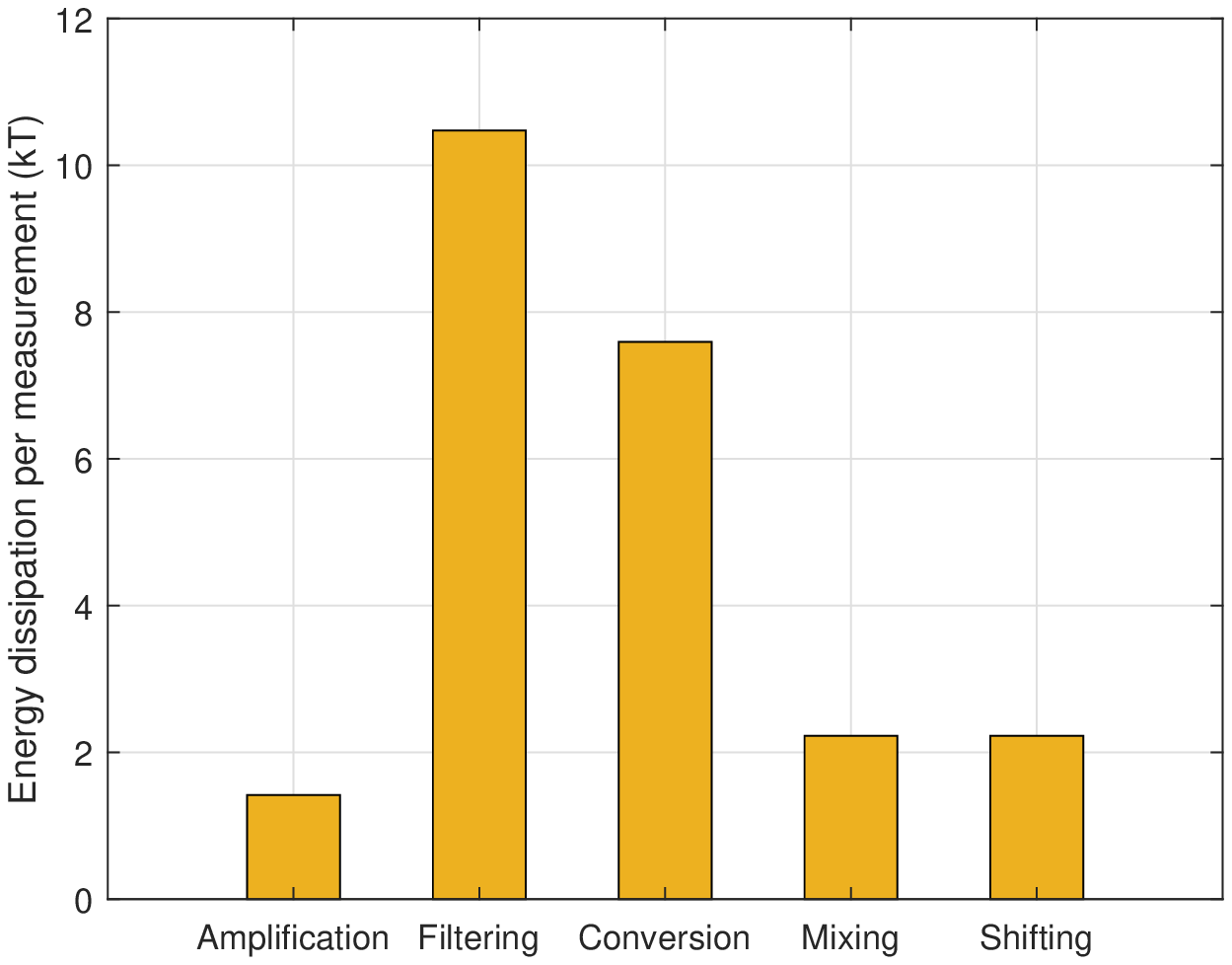}
\begin{quote}
\small Fig. 7. Energy dissipation of the information transmission in the measurement-erasure cycle.
\end{quote}
\end{figure}

\begin{figure}[!htb]
\centering
\includegraphics[width=3in]{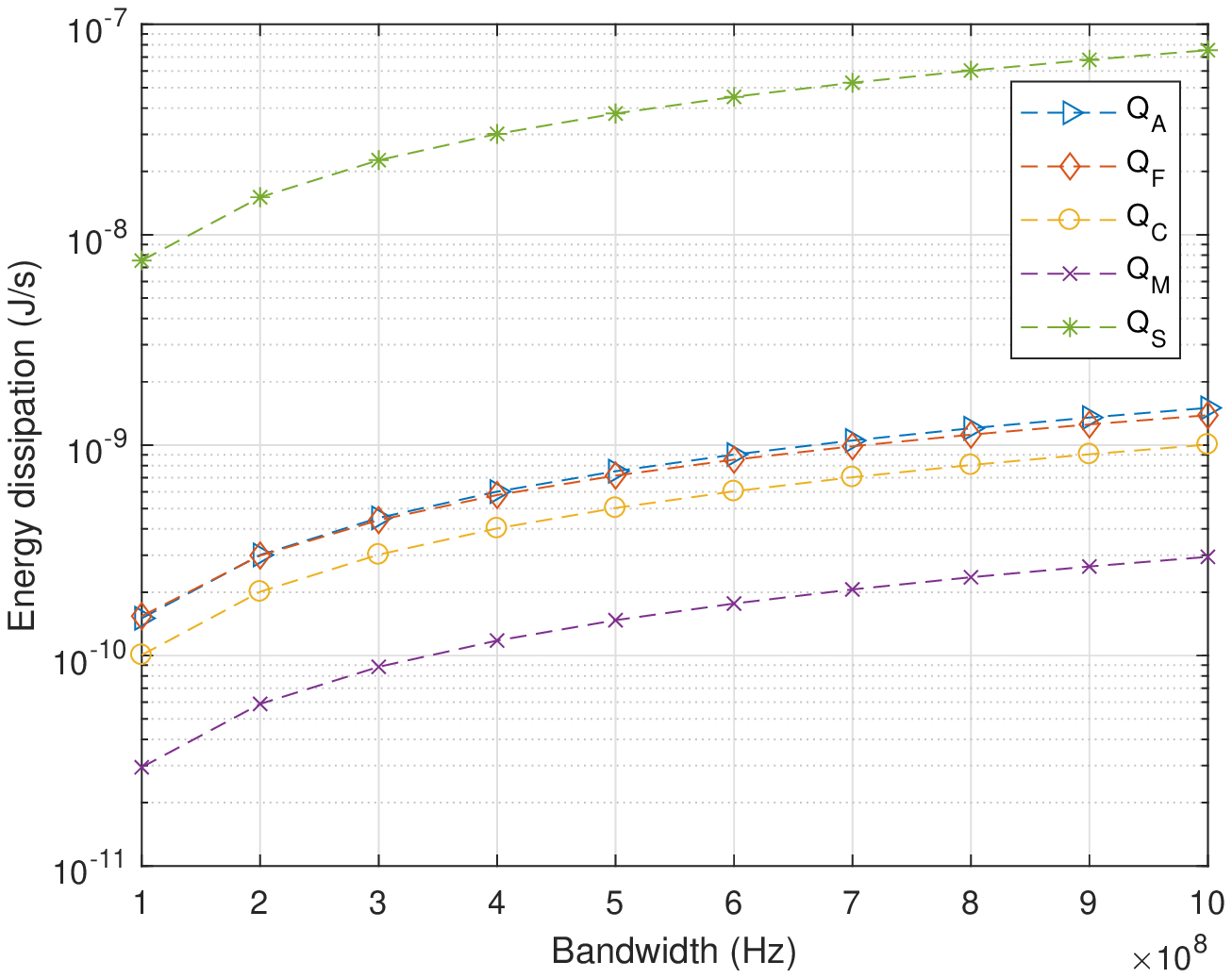}
\begin{quote}
\small Fig. 8. Energy dissipation of the information transmission with respect to bandwidth with realistic circuit structure.
\end{quote}
\end{figure}

For information transmission with realistic circuit implementation, the relationship between the energy dissipation and the bandwidth is illustrated in Fig. 8. When the bandwidth increases, the energy dissipation of modulations of amplification, filtering, analog-to-digital conversion, frequency mixing and phase shifting are increased. When the bandwidth is fixed, the energy dissipation of the phase shifting modulation is the largest, while the energy dissipation of the frequency mixing modulation is the least.

The minimum energy dissipation, i.e., the energy dissipation without the residual dissipation, and the typical energy dissipation adopting current technologies are compared for mobile communication systems in Fig. 9. Fig. 9(a) illustrates the comparison between the minimum energy dissipation and practical energy dissipation for the information processing of mobile communication systems. Simulation results show that the minimal energy dissipation limit is about 3 orders of magnitude away from the practical energy dissipation for the linear processing modulation and channel estimation modulation, which is consistent with the measured results in \cite{luca2015}. The minimal energy dissipation limit is about 8 orders of magnitude away from the practical energy dissipation for the channel coding modulation of mobile communication systems. Based on the result of Fig. 9(b), the gap between the practical energy dissipation and the theoretical minimal energy dissipation limit is more than 7 orders of magnitude for the information transmission.
\begin{figure}[!htb]
\centering
\includegraphics[width=3in]{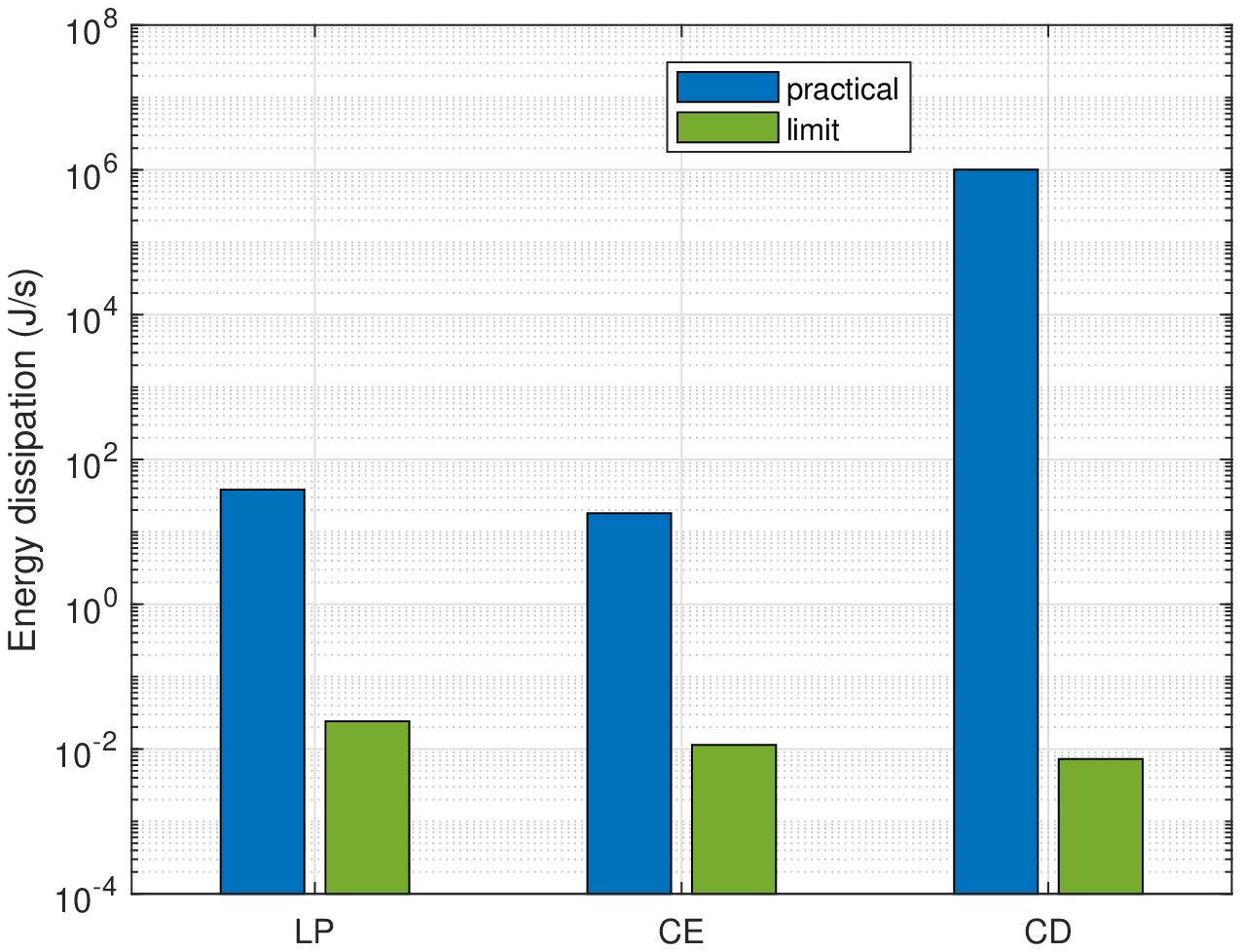}
\begin{quote}
\small Fig. 9(a). Comparison of the energy dissipation of information processing between the theoretical minimal limit and practical wireless communication systems.
\end{quote}
\end{figure}
\begin{figure}[!htb]
\centering
\includegraphics[width=3in]{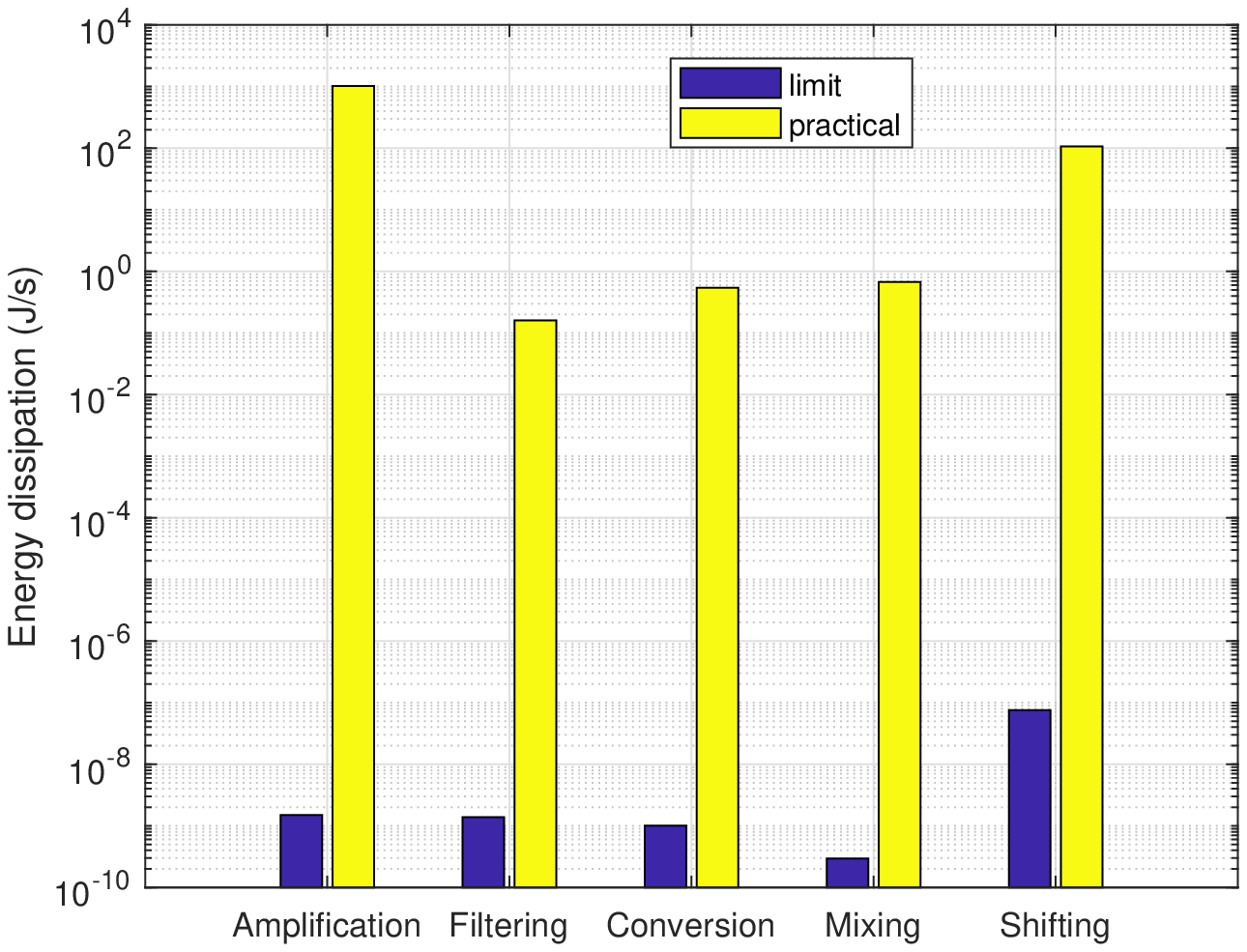}
\begin{quote}
\small Fig. 9(b). Comparison of the energy dissipation of information transmission between the theoretical minimal limit and practical wireless communication systems.
\end{quote}
\end{figure}

\section{Conclusions}
\label{sec6}
In this paper, we propose a new entropy-based energy dissipation model for mobile communication systems. Considering different thermodynamic properties of operations in mobile communication systems, the energy dissipation of eight typical modulations are derived for mobile communication systems. Moreover, the theoretical minimal energy dissipation of information processing and information transmission is proposed for mobile communication systems. Simulation results show that there is a gap of three orders of magnitude between the practical energy dissipation of information processing and the minimal energy dissipation limit of information processing for mobile communication systems. The gap between the practical energy dissipation of information transmission and the minimal energy dissipation limit of information transmission is more than seven orders of magnitude for mobile communication systems. In the future work, the proposed energy dissipation model can provide guidelines for optimizing the energy efficiency of mobile communication systems.


\begin{thebibliography}{1}
\bibitem{ge2017}
X. Ge, et al., ``Energy efficiency challenges of 5G small cell networks,'' in {\em IEEE Commun. Mag.},vol. 55, no. 5, pp. 184--191, 2017.

\bibitem{conte2019}
T. Conte et al., ``Thermodynamic Computing,'' 2019,  {\em arXiv: 1911.01968}. [Online]. Available: https://arxiv.org/abs/1911.01968

\bibitem{lopez2013}
D. Lopez-Perez, X. Chu, A. V. Vasilakos, and H. Claussen, ``Power minimization based resource allocation for interference mitigation in OFDMA femtocell networks,'' in {\em IEEE j. sel. areas commun.}, vol. 32, no. 2, pp. 333--344, 2014.

\bibitem{merti2016}
P. Mertikopoulos and E. V. Belmega, ``Learning to be green: Robust energy efficiency maximization in dynamic MIMO–OFDM systems,'' in {\em IEEE j. sel. areas commun.}, vol. 34, no. 4, pp. 743--757, 2016.

\bibitem{heath2016}
R. W. Heath, N. Gonzalez-Prelcic, S. Rangan, W. Roh, and A. M. Sayeed, ``An overview of signal processing techniques for millimeter wave MIMO systems,'' in {\em IEEE J. Sel. Top. Signal Process.}, vol. 10, no. 3, pp. 436--453, 2016.

\bibitem{ge2018}
X. Ge, Y. Sun, H. Gharavi, and J. Thompson, ``Joint optimization of computation and communication power in multi-user massive Mimo systems,'' in {\em IEEE Trans. Wirel. Commun.}, vol. 17, no. 6, pp. 4051--4063, 2018.

\bibitem{zhang2020}
S. Zhang and R. Zhang, ``Capacity characterization for intelligent reflecting surface aided MIMO communication,'' in {\em IEEE j. sel. areas commun.}, vol. 38, no. 8, pp. 1823--1838, 2020.

\bibitem{yangz2022}
Z. Yang et al., ``Energy-efficient wireless communications with distributed reconfigurable intelligent surfaces,'' in {\em IEEE Trans. Wirel. Commun.}, vol. 21, no. 1, pp. 665--679, 2022.

\bibitem{miao2010}
G. Miao, N. Himayat, and G. Y. Li, ``Energy-efficient link adaptation in frequency-selective channels,'' in {\em IEEE trans. commun}., vol. 58, no. 2, pp. 545--554, 2010.

\bibitem{visser2013}
H. J. Visser and R. J. M. Vullers, ``RF energy harvesting and transport for wireless sensor network applications: Principles and requirements,'' in {\em Proc. IEEE Inst. Electr. Electron. Eng.}, vol. 101, no. 6, pp. 1410--1423, 2013.

\bibitem{perera2017}
T. D. Perera, ``Simultaneous wireless information and power transfer (SWIPT): Recent advances and future challenges,'' in {\em IEEE Communications Surveys and Tutorials}, vol. 20, pp. 264--302, 2017.

\bibitem{wagih2021}
M. Wagih, G. S. Hilton, A. S. Weddell, and S. Beeby, ``Dual-band dual-mode textile antenna/rectenna for simultaneous wireless information and power transfer (SWIPT),'' in {\em IEEE Trans. Antennas Propag.}, vol. 69, no. 10, pp. 6322--6332, 2021.

\bibitem{zhirnov2014}
V. Zhirnov, R. Cavin, and L. Gammaitoni, ICT-Energy-Concepts Towards Zero-Power Information and Communication Technology. IntechOpen. 2014.

\bibitem{landauer1961}
R. Landauer, ``Irreversibility and heat generation in the computing process,'' in {\em IBM J. Res. Dev.}, vol. 5, no. 3, pp. 183--191, 1961.

\bibitem{berut2012}
A. Bérut et al., ``Experimental verification of Landauer’s principle linking information and thermodynamics,'' in {\em Nature}, vol. 483, no. 7388, pp. 187--189, 2012.

\bibitem{luca2015}
G. Luca, D. Chiuchiú, and M. Madami, ``Towards zero-power ICT,'' in {\em Nanotechnology}, vol. 26, no. 22, 2015.

\bibitem{crooks}
G. E. Crooks, ``Entropy production fluctuation theorem and the nonequilibrium work relation for free energy differences,'' in {\em Phys. Rev. E Stat. Phys. Plasmas Fluids Relat. Interdiscip. Topics}, vol. 60, no. 3, pp. 2721--2726, 1999.

\bibitem{jarzynski}
C. Jarzynski, ``Nonequilibrium equality for free energy differences,'' in {\em Phys. Rev. Lett.}, vol. 78, no. 14, pp. 2690--2693, 1997.

\bibitem{diamantini2016}
M. C. Diamantini, L. Gammaitoni, and C. A. Trugenberger, ``Landauer bound for analog computing systems,'' in {\em Phys. Rev. E.}, vol. 94, no. 1, p. 012139, 2016.

\bibitem{boyd1}
A. B. Boyd, D. Mandal, and J. P. Crutchfield, ``Identifying functional thermodynamics in autonomous Maxwellian ratchets,'' in {\em New J. Phys.}, vol. 18, no. 2, p. 023049, 2016.

\bibitem{boyd2}
A. B. Boyd, D. Mandal, and J. P. Crutchfield, ``Thermodynamics of modularity: Structural costs beyond the Landauer bound,'' in {\em Physical Review X}, vol. 8, no. 3, 2018.

\bibitem{boyd3}
A. B. Boyd, A. Patra, C. Jarzynski, and J. P. Crutchfield, ``Shortcuts to thermodynamic computing: The cost of fast and faithful information processing,'' in {\em J. Stat. Phys.},  vol. 151, no. 2, pp. 174--202, 2022.

\bibitem{aghamohammadi2017}
C. Aghamohammadi and J. P. Crutchfield, ``Thermodynamics of random number generation,'' in {\em Phys. Rev. E.}, vol. 95, no. 6, p. 062139, 2017.

\bibitem{kolchinsky2020}
A. Kolchinsky and D. H. Wolpert, ``Thermodynamic costs of Turing machines,'' in {\em Phys. Rev. Research}, vol. 2, no. 3, 2020.

\bibitem{parrondo2015}
J. M. R. Parrondo, J. M. Horowitz, and T. Sagawa, ``Thermodynamics of information,'' in {\em Nat. Phys.}, vol. 11, no. 2, pp. 131--139, 2015.

\bibitem{seifert2012}
U. Seifert, ``Stochastic thermodynamics, fluctuation theorems and molecular machines Rep,'' in {\em Rep. Prog. Phys.}, vol. 75, 2012.

\bibitem{peliti2021}
L. Peliti and S. Pigolotti, Stochastic Thermodynamics: An Introduction. Princeton University Press, 2021.

\bibitem{bennett2003}
C. H. Bennett, ``Notes on Landauer’s principle, reversible computation, and Maxwell’s Demon,'' in {\em Stud. Hist. Philos. Sci. B Stud. Hist. Philos. Modern Phys.}, vol. 34, no. 3, pp. 501--510, 2003.

\bibitem{ge2022}
X. Ge and L. Yan, “Information Thermodynamics Communications,'' in {\em IEEE Wirel. Commun.}, pp. 1--14, 2022.

\bibitem{wolpert2020}
D. H. Wolpert and A. Kolchinsky, ``Thermodynamics of computing with circuits,'' in {\em New J. Phys.}, vol. 22, no. 6, p. 063047, 2020.

\bibitem{bird2014}
J. Bird, Engineering Mathematics, 8th ed. London, England: Routledge, 2017.

\bibitem{mohammed2014}
S. K. Mohammed, ``Impact of transceiver power consumption on the energy efficiency of zero-forcing detector in massive MIMO systems,'' in {\em IEEE trans. commun.}, vol. 62, no. 11, pp. 3874--3890, 2014.

\bibitem{boyd2004}
S. P. Stephen and L. Boyd, Convex optimization. Cambridge university press, 2004.

\bibitem{golub2013}
G. H. Golub and C. F. Van Loan, Matrix Computations, 4th ed. Baltimore, MD: Johns Hopkins University Press, 2013.

\bibitem{bjornson2015}
E. Bjornson, L. Sanguinetti, J. Hoydis, and M. Debbah, ``Optimal design of energy-efficient multi-user MIMO systems: Is massive MIMO the answer?'' in {\em IEEE Trans. Wirel. Commun.}, vol. 14, no. 6, pp. 3059--3075, 2015.

\bibitem{masera2005}
G. Masera, F. Quaglio, and F. Vacca, ``Finite precision implementation of LDPC decoders,'' in {\em IEE Proc. - Commun.}, vol. 152, no. 6, pp. 1098--1102, 2005.

\bibitem{desset2003}
C. Desset and A. Fort, ``Selection of channel coding for low-power wireless systems,'' in {\em The 57th IEEE Semiannual Vehicular Technology Conference}, Vol. 3, pp. 1920--1924, 2003.

\bibitem{kong2019}
L. Kong, H. Pan, X. Li, S. Ma, Q. Xu, and K. Zhou, ``An information entropy-based modeling method for the measurement system,'' in {\em Entropy}, vol. 21, no. 7, p. 691, 2019.

\bibitem{sripad}
A. Sripad and D. Snyder, ``A necessary and sufficient condition for quantization errors to be uniform and white,'' in {\em IEEE Trans. Acoust.}, vol. 25, no. 5, pp. 442--448, 1977.

\bibitem{yang2022}
J. Yang, X. Ge, and Y. Li, ``Principle of computation power optimization in millimeter wave massive MIMO systems,'' in {\em IEEE Trans. Mob. Comput.}, vol. 21, no. 8, pp. 2955--2966, 2022.

\bibitem{yang2019}
X. Yang, et al., ``Hardware-constrained millimeter-wave systems for 5G: Challenges, opportunities, and solutions,'' in {\em IEEE Commun. Mag.}, vol. 57, no. 1, pp. 44--50, 2019.

\end{thebibliography}
\end{document}